\documentstyle[psfig]{texas}
\makeatletter
 \def\@biblabel#1{}
\makeatother

\def\AJ{{\em Astron. J.} }
\def\AA{{\em Astron. Astrophys.} }

\def\AAS{{\em Astron. Astrophys. Suppl. Ser.} }
\def\ARAA{{\em Ann. Rev. Astron. Astrophys.} }
\def\APJ{{\em Astrophys. J.} }
\def\APJL{{\em Astrophys. J. Lett.} }
\def\APJS{{\em Astrophys. J. Suppl.} }
\def\MN{{\em Mon. Not. R. Astr. Soc.} }
\def\MNRAS{{\em Mon. Not. R. Astr. Soc.} }
\def\PASP{{\em Publ. Astr. Soc. Pacific\/} }

\def\HST{{\sl HST\/ }}
\def\la{\mathrel{\hbox{\rlap{\hbox{\lower4pt\hbox{$\sim$}}}\hbox{$<$}}}}
\def\ga{\mathrel{\hbox{\rlap{\hbox{\lower4pt\hbox{$\sim$}}}\hbox{$>$}}}}
\def\kms{\nobreak\mbox{$\;$km\,s$^{-1}$}}
\def\Mpc{\nobreak\mbox{$\;$Mpc}}
\def\mag{\nobreak\mbox{$^{\rm m}$}\!\!\!\!.\;}
\def\tablefontsize{\small\rm}

\begin{document}

\noindent
\parbox{10cm}{\small 19th Texas Symposium on Relativistic Astrophysics
  and Cosmology, Paris December 14--18, 1998.}

\title{THE DISTANCE OF THE VIRGO CLUSTER}
\author{G.A. Tammann$^{1}$, A. Sandage$^{2}$,  \&  B. Reindl$^{1}$}
\address{(1) Astronomisches Institut der Universit{\"a}t Basel, \\
       Venusstr.~7, CH-4102 Binningen, Switzerland \\ }
\address{(2) The Observatories of the Carnegie Institution of Washington, \\ 
         813 Santa Barbara Street, Pasadena, CA 91191--1292, USA}

\begin{abstract}
Six different distance determination methods of Virgo cluster members
yield a mean distance modulus of $(m-M)^0_{\rm Virgo}=31.60\pm0.09$
($20.9\pm0.9\Mpc$). This value can be carried out to $\ga 10\,000\kms$
by means of 31 clusters whose distances are well known {\em
  relative\/} to the Virgo cluster. They yield $H_0=56\pm4$ (random
error) $\pm6$ (systematic error), independent of local streaming motions.
\end{abstract}

\section{Introduction} 
The Virgo cluster is a fundamental milestone for the determination of
$H_0$ because it is the nearest reasonably rich cluster which is
tightly tied into the large-scale expansion field through excellent
{\em relative\/} distances to more distant clusters, circumventing
thus effects of local streaming velocities.

   The distance of the Virgo cluster has long been controversial
mainly for three reasons. (1)~The cluster which spans $\ga 15^{\circ}$
in the sky has a considerable depth effect making even good distance
determinations of {\em few\/} member galaxies vulnerable to small-number
effects. This becomes particularly precarious if the selection
criteria of the few members are themselves distance-dependent. 
(2)~Distance indicators with non-negligible intrinsic scatter lead
always to too small distances if they are applied to only the
brightest cluster members (Malmquist bias). This has made Virgo
cluster distances useless especially from the 21cm-line width
(Tully-Fisher) method until a very deep Virgo cluster catalogue became
available (Binggeli, Sandage, \& Tammann 1985).
The catalogue is {\em complete\/} as to normal E and spiral galaxies 
because it goes below the cutoff magnitude of these objects. 
(3)~Undue weight has been given to distance indicators in the past 
which had never been tested in the relevant distance range. In the 
case of the bright tail of the luminosity function of planetary 
nebulae the ineffectiveness as distance indicator is now explained by
the dependence on the sample (i.e. galaxy) size (Bottinelli
et~al. 1991; Tammann 1993; M{\'e}ndez et~al. 1993; Soffner et~al. 1996).
The reasons why surface brightness fluctuations of E galaxies fail to 
provide useful absolute distances beyond $10\;$Mpc are less clear. The 
applicability of the method to dwarf ellipticals is presently 
investigated (Jerjen, Freeman, \& Binggeli 1998). 

   With these difficulties in mind six different distance 
determinations of the Virgo cluster are discussed in the following. 

\section{The Virgo cluster distance from Cepheids} 
\label{sec:2}
Cepheids are presently, through their period-luminosity (PL) 
relation, the most reliable and least controversial distance 
indicators. The slope and the zeropoint of the PL relation is taken 
from the very well-observed Cepheids in the Large Magellanic Cloud 
(LMC), whose distance modulus is adopted to be $(m-M) = 18.50$ 
(Madore \& Freedman 1991). 

   An old PL relation calibrated by Galactic Cepheids in open clusters,
and now vindicated by Hipparcos data (Sandage \& Tammann 1998), gave
$(m-M)_{\rm LMC}=18.59$ (Sandage \& Tammann 1968, 1971). Hipparcos
data combined with more modern Cepheid data give an even somewhat
higher modulus (Feast \& Catchpole 1997). Reviews of Cepheid distances
(Federspiel, Tammann, \& Sandage 1998; Gratton 1999) cluster around
$18.56\pm0.05$, -- a value in perfect agreement with the purely
geometrical distance determination of SN\,1987A ($18.58\pm0.05$;
Gilmozzi \& Panagia 1999). In his excellent review Gratton (1999)
concludes from the rich literature on RR\,Lyr stars that $(m-M)_{\rm
  LMC}=18.54\pm0.12$. He also discusses five distance determination
methods which give lower moduli by $0.1-0.2\;$mag, but they are still
at a more experimental stage. -- There is therefore emerging evidence
that the adopted LMC modulus of 18.50 is too small by $\sim\!0.06$ or
even $0.12\;$mag (Feast 1999) 
and that all Cepheid distances in the following should be increased by this
amount.

   There has been much debate about the possibility that the PL 
relation of Cepheids depends on metallicity. Direct 
{\it observational\/} evidence for a (very) weak metallicity 
dependence comes from the fact that the metal-rich 
Galactic Cepheids give perfectly reasonable distances for the 
moderately metal-poor LMC Cepheids and the really metal-poor SMC 
Cepheids and, still more importantly, that their relative distances  
are wavelength-independent (Di Benedetto 1997; cf. Tammann 1997).  
-- Much progress has been made on the theoretical front. Saio 
\& Gautschy (1998) and Baraffe et~al. (1998) have evolved Cepheids 
through the different crossings of the instability strip and have 
investigated the pulsational behavior at any point. The resulting 
(highly metal-insensitive) PL relations in bolometric light have been 
transformed into PL relations at different wavelengths by means of 
detailed atmospheric models; the conclusion is that any metallicity 
dependence of the PL relations is negligible (Sandage, Bell, \& 
Tripicco 1998; Alibert et~al. 1999; cf. however Bono, Marconi, \& 
Stellingwerf 1998, who strongly depend on the treatment of stellar 
convection).  

   Most extragalactic Cepheid distances are now due to \HST 
observations. The reduction of these observations is by no means 
simple. The photometric zeropoint, the linearity over the field, 
crowding, and cosmic rays raise technical problems. The quality of the 
derived distances depends further on (variable) internal absorption 
and the number of available Cepheids in view of the finite width of 
the instability strip. (An attempt to beat the latter problem by using 
a PL-color relation is invalid because the underlying assumption of 
constant slope of the constant-period lines is unrealistic; cf. Saio 
\& Gautschy 1998). Typical errors of individual Cepheid distances from  
\HST are therefore $\pm0.2\,$mag (10\% in distance).

There are now three bona fide cluster members and two outlying members
(cf. Binggeli, Popescu, \& Tammann 1993) with Cepheid distances from \HST 
(Table~\ref{tab:Virgo_Cepheid_distances}; cf. Freedman et~al. 1998).
%
\begin{table}[t]
\tablefontsize
\caption{The Virgo cluster members with Cepheid distances}
\label{tab:Virgo_Cepheid_distances}
\begin{center}
\begin{minipage}{0.8\textwidth}
\begin{tabular}{llll}
\noalign{\smallskip}
\hline
\noalign{\smallskip}
  Galaxy & $(m-M)_{\rm Cepheids}$ & Remarks & $(m-M)_{\rm TF}$ \\
\noalign{\smallskip}
\hline
\noalign{\smallskip}
  NGC\,4321$^*$ & $31.04\pm0.21$ & highly resolved & $31.21\pm0.40$ \\
  NGC\,4496A$^{*\,*}$
            & $31.13\pm0.10$ & highly resolved & $30.67\pm0.40$ \\
  NGC\,4536$^{*\,*}$
            & $31.10\pm0.13$ & highly resolved & $30.72\pm0.40$ \\
  NGC\,4571 & $30.87\pm0.15$ & extremely resolved & $31.75\pm0.40$ \\
  NGC\,4639 & $32.03\pm0.23$ & poorly resolved & $32.53\pm0.40$ \\
\noalign{\smallskip}
\hline
\noalign{\smallskip}
\end{tabular}

{\footnotesize
\hspace*{5pt}$^*$~From a re-analysis of the {\sl HST\/} observations 
Narasimha \& Mazumdar (1998) obtained $(m-M)=31.55\pm0.28$. \\ 
$^{*\,*}$~In the W-cloud outside the confidence  boundaries of the
Virgo cluster (cf. Federspiel et~al. 1998).
}
\end{minipage}
\end{center}
\end{table}
The wide range of their distance moduli, corresponding to $14.9$ to 
$25.5\;$Mpc, reveals the important depth effect of the cluster. The 
first four galaxies in Table~\ref{tab:Virgo_Cepheid_distances} have 
been chosen from the atlas of Sandage \& Bedke (1988)
because they are highly resolved and
seemed easy as to their Cepheids. They are therefore {\em expected\/} 
to lie on the near side of the cluster. In contrast NGC\,4639 has been 
chosen as parent to SN\,1990N and hence independently of its distance
(Saha et~al. 1997); correspondingly this distance is expected to be
statistically more representative. 
A straight mean of the distances in
Table~\ref{tab:Virgo_Cepheid_distances} is therefore likely to be an
underestimate. Indeed the mean Tully-Fisher (TF) distance modulus of
the five galaxies is $0\mag2$ (corresponding to $10\%$ in distance)
{\em smaller\/} than the mean distance of a complete and fair sample
of TF distances (Federspiel et~al. 1998).
-- It should be noted that NGC\,4639 with the largest distance in
Table~\ref{tab:Virgo_Cepheid_distances} has a recession velocity of
only $v_0=820\kms$, i.e. {\em less\/} than the mean cluster
velocity of $v_0=920\kms$, and that it can therefore not be
assigned to the background. In fact the redshift distribution in
the Virgo cluster area shows a pronounced gap behind the cluster
minimizing the danger of background contamination (Binggeli
et~al. 1993).

   B{\"o}hringer et~al. (1997) have proposed that the Cepheid
distances of the spiral galaxies NGC\,4501 and NGC\,4548 would be
significant because these galaxies are spatially close to the Virgo
cluster center on the basis of their being stripped by the X-ray
intracluster gas. In the case of NGC\,4548 there are some doubts
because it is, like NGC\,4571, exceptionally well resolved (Sandage \&
Bedke 1988).
The resolution of NGC\,4501 is about intermediate between the two
last-mentioned galaxies and the poorly resolved NGC\,4639. A first
rough distance of NGC\,4501 is provided by the TF method
(Section~\ref{sec:4}) which gives $(m-M)=31.5\pm0.4$ in good agreement
with the preferred mean Virgo cluster distance. However, the inherent
errors of the TF method if applied to individual galaxies prevent a
stringent test. NGC\,4501 is therefore an interesting candidate for
Cepheid observations.

   A preliminary Cepheid distance of the Virgo cluster is obtained by 
taking the Cepheid distance of the Leo group of $(m-M)=30.20\pm0.12$, 
based now on three galaxies with Cepheids from {\sl HST} (Saha 
et~al. 1999), and to step up this value by the modulus difference of
$\Delta(m-M)=1.25\pm0.13$  (Tammann \& Federspiel 1997)
between the Leo group and the Virgo 
cluster. The result is $(m-M)_{\rm Virgo}=31.45\pm0.21$, a value which
is well embraced by the individual Cepheid distances in 
Table~\ref{tab:Virgo_Cepheid_distances}.

\section{The Virgo cluster distance from Supernovae type Ia} 
\label{sec:3}
Blue SNe\,Ia (i.e. $B_{\max}-V_{\max} \le 0.20$) at maximum light are 
nearly perfect standard candles. After small corrections for second 
parameters (decline rate and color) their scatter about the mean 
Hubble line amounts to only $0\mag12$ for $v > 10\,000\kms$ (Parodi 
et~al. 1999). The equally corrected absolute 
peak magnitude of $M_{\rm B}^{\rm corr}=-19.44\pm0.04$, based on 
seven\,(!) SNe\,Ia with known Cepheid distances (Parodi 
et~al. 1999), is secure. The value is also in
perfect agreement with present theoretical models (Branch 1998).

   Four blue SNe\,Ia with complete photometry are known in the Virgo 
cluster (Table~\ref{tab:SNeIa_Virgo}). The photometric data from the 
Tololo/Calan survey are compiled by Parodi et~al. (1999).
The magnitudes, corrected for second parameters ($\Delta m_{15}$ and
color), are calculated from equation~(30) in Parodi et~al. (1999).

%
\begin{table}[t]
\tablefontsize
\begin{center}
\caption{Blue SNe\,Ia with good photometry in the Virgo cluster}
\label{tab:SNeIa_Virgo}
\begin{tabular}{llrrrl}
\noalign{\smallskip}
\hline
\noalign{\smallskip}
  SN & Galaxy & $m_{\rm B}$ & $m_{\rm V}$ &  $\Delta m_{15}$ &
  $m_{\rm B}^{\rm corr}$ \\
\noalign{\smallskip}
\hline
\noalign{\smallskip}
1981B & NGC\,4536 & 12.04 & 11.96 & 1.10 & 11.89 \\
1984A & NGC\,4419 & 12.45 & 12.26 & 1.20 & 12.06 \\  
1990N & NGC\,4639 & 12.76 & 12.70 & 1.03 & 12.68 \\
1994D & NGC\,4526 & 11.86 & 11.87 & 1.27 & 11.81 \\
1960F & NGC\,4496A & 11.60 & 11.54 & \multicolumn{1}{c}{---} & --- \\ 
\noalign{\smallskip}
\hline
\noalign{\smallskip}
 \multicolumn{2}{c}{mean:} & & & & $12.11\pm0.20$ \\
\noalign{\smallskip}
\hline
\noalign{\smallskip}
\multicolumn{6}{l}{The spectroscopically unusual SN\,1991T is omitted.}
\end{tabular}
\end{center}
\end{table}

   With the above calibration for $M_{\rm B}^{\rm corr}$ and the mean
value of $m_{\rm B}^{\rm corr}$ in Table~\ref{tab:SNeIa_Virgo} the
cluster modulus becomes $(m-M)_{\rm Virgo}=31.55\pm0.20$.

   The error of the result is dominated by the depth effect of the
cluster. Additional SNe\,Ia in the cluster will improve the result
considerably. Its present advantage over the Cepheid distance of the
cluster is that the four SNe\,Ia have been discovered independently of
their position within the cluster.

   To increase the sample one additional blue SNe\,Ia in the Virgo
cluster may be added which, however, has no known decline rate $\Delta
m_{15}$ (cf. Table~\ref{tab:SNeIa_Virgo}). But since the Virgo SNe\,Ia
and the calibrating SNe\,Ia (except one) lie in spirals the
second-parameter correction cancels in first approximation. {\em
  Eight\/} SNe\,Ia with Cepheid distances (including SN\,1960F
without $\Delta m_{15}$) give a straight calibration of $M_{\rm
  B}=M_{\rm V}=-19.48\pm0.04$ (Saha et~al. 1999), 
while the five Virgo SNe\,Ia in Table~\ref{tab:SNeIa_Virgo} have a
mean (uncorrected) apparent magnitude of $m_{\rm B}=12.14\pm0.21$,
$m_{\rm V}=12.07\pm0.20$. From this follows a mean Virgo cluster
modulus of $(m-M)_{\rm Virgo}=31.59\pm0.15$ which is essentially
undistinguishable from the result of the four corrected SNe\,Ia.

\section {The Virgo cluster distance from 21cm-line widths} 
\label{sec:4}
The method using 21~cm line-widths, the so-called Tully-Fisher (TF)
relation, has been applied many times but with variable 
success. Widely divergent values are in the literature which in some 
cases favor the short distance scale (e.g. Pierce \& Tully 1992 
with $m - M = 30.9$) and in others the long scale 
(Kraan-Korteweg et al.~1988 with $m - M = 31.6$; Fouqu{\'e} et~al. 1990 
with the same value if corrected to the modern local calibrators; 
Federspiel et~al. 1998). 

  It has been shown (Federspiel et al.~1994; Sandage, 
Tammann, \& Federspiel 1995) that the reasons for small 
values of $(m - M)$ for Virgo (the short scale) using TF are two; (1) 
use of incorrectly small distances to the local calibrators in earlier  
papers by proponents of the short scale, and (2) neglect of the  
disastrous effect of the Teerikorpi (1987, 1990) 
cluster incompleteness bias. It 
can be shown that this bias produces errors in the modulus up to 1 mag 
depending on how far one has sampled into the cluster luminosity 
function {\em regardless how the sample is chosen\/}, if only the sample 
is cut by apparent magnitude. The modulus error is a strong function of the 
fraction of the luminosity function that remains unsampled 
(Kraan-Korteweg et al. 1988; Sandage et al.~1995).

     The calibration of the TF relation has been much 
improved by the advent of Cepheid distances with HST. There are 
now 18 Cepheid distances available for spirals suitable for the 
calibration. Detailed data with complete references to the 
extensive literature are given elsewhere (Tammann \& Federspiel 
1997; Federspiel et~al. 1998) and are not repeated here.  

     The most recent applications of the TF relation on the Virgo
cluster (Schr{\"o}der 1996; Tammann \& Federspiel 1997; Federspiel
et~al. 1998) 
use a {\em complete\/} sample of Virgo cluster spirals. Rigid 
criteria have been invoked in the selection of members within the
cluster boundaries defined by counts, redshifts, and X-ray contours. Several 
subtleties, not seen in earlier studies, have been found. These 
include a variation of the derived modulus on the 
wavelength of the observations (covering UBVRI), and a 
correlation of the derived modulus on the degree of hydrogen 
depletion for the spirals. With this in mind Federspiel et~al. (1998) 
have derived from a complete sample of 49 sufficiently inclined 
spirals  
\begin{displaymath}
  (m-M)_{\rm Virgo}=31.58\pm0.24.
\end{displaymath}

\section{The Virgo Cluster distance from Globular Clusters}
\label{sec:5}
Extragalactic GCs, discovered in M\,31 by Hubble (1932), took a
role in distance determinations when Racine (1968) proposed the
bright end of the globular cluster luminosity function (GCLF) to be
used as a ``standard candle''. First 
applications of this tool provided reasonable distances to M\,87
(Sandage 1968, de Vaucouleurs 1970), yet it was soon realized that the
results were sensitive to the GC population size, and that the
luminosity $M^{\ast}$ of the turnover point of the bell-shaped GCLF is
a much more stable standard candle. This required, however, that one
had to sample at least four magnitudes into the GCLF which became
feasible only with the advent of CCDs. The first application of the
new method to a giant E galaxy (M\,87; van den Bergh et~al. 1985)
was followed by many papers such that $m^{\ast}$ magnitudes are now
available for about two dozen full-size galaxies 
(for reviews e.g. Harris 1991; Whitmore 1997; Tammann \& Sandage 1999).

   The absolute magnitude $M^{\ast}$ of the peak of the globular
cluster luminosity function (GCLF), approximated by a Gaussian, can
be calibrated independently in the Galaxy and M\,31 through RR\,Lyr
stars and Cepheids, respectively. They yield, in perfect agreement,
$M^{\ast}_{\rm B} = -6.93\pm0.08$ and  $M^{\ast}_{\rm V} =
-7.62\pm0.08$ (Sandage \& Tammann 1995; Tammann \& Sandage 1999).
Remaining differences between different authors of the luminosity
calibration of RR\,Lyr stars are vanishingly small for the mean
metallicity of the Galactic GCs of [Fe/H]$=-1.35$, because different
adopted luminosity-metallicity relations meet for this value very
closely at $M_{\rm V}({\rm RR})=0\mag54$. The calibration of
$M^{\ast}$, independently confirmed by the M\,31 Cepheids, is
therefore uncontroversial. 

   Different values of $m^{\ast}_{\rm B}$ and $m^{\ast}_{\rm V}$ of bona fide
members of the Virgo cluster (cf. Binggeli et~al. 1985) are
compiled in Table~\ref{tab:Virgo_GC}. The values have been corrected for the
small and variable Galactic absorption according to Burstein \& Heiles
(1984). The  $g$ magnitudes of Cohen (1988) and the $R$ magnitudes of
Ajhar et~al. (1994) were transformed into $V$ magnitudes following
Whitmore (1997). No (precarious) attempt was made to transfer
$m^{\ast}_{\rm B}$ into $m^{\ast}_{\rm V}$. 

%
\begin{table}[t]
\tablefontsize
\begin{center}
\caption{Virgo cluster members with known turnover magnitude
  $m^{\ast}$ of the GCLF}
\label{tab:Virgo_GC}
\begin{minipage}{0.9\textwidth}
\begin{tabular}{lllrc}
\noalign{\smallskip}
\hline
\noalign{\smallskip}
  Galaxy & \multicolumn{1}{c}{$m^{\ast}_{\rm B}$} &
  \multicolumn{1}{c}{$m^{\ast}_{\rm V}$}   & 
  \multicolumn{1}{c}{$m^{\ast}_{\rm B}$ - $m^{\ast}_{\rm V}$}  & Source \\
\noalign{\smallskip}
\hline
\noalign{\smallskip}
 NGC\,4365 & $25.18\pm0.16(2)$ & $24.47\pm0.21(1)$ & $0.71\pm0.26$ & (1)\\
 NGC\,4374 &                   & $24.12\pm0.30(1)$ &               & (2)\\
 NGC\,4406 &                   & $24.25\pm0.30(1)$ &               & (2)\\
 NGC\,4472 & $24.70\pm0.11(1)$ & $23.85\pm0.21(2)$ & $0.85\pm0.24$ & (3)\\
 NGC\,4486 & $24.82\pm0.11(2)$ & $23.74\pm0.06(5)$ & $1.08\pm0.13$ & (4)\\
 NGC\,4552 &                   & $23.70\pm0.30(1)$ &               & (2)\\
 NGC\,4636 &                   & $24.18\pm0.20(1)$ &               & (5)\\
 NGC\,4649 & $24.65\pm0.14(1)$ &                   &               & (6)\\ 
\noalign{\smallskip}
\hline
\noalign{\smallskip}
 straight mean: & $24.84\pm0.12$ & $24.03\pm0.10$ & & \\
 $(m-M)$:       & $31.77\pm0.14$ & $31.65\pm0.13$ & & \\
\noalign{\smallskip}
 $\Rightarrow$  & \multicolumn{2}{l}{$(m-M)=31.70\pm0.10$} & (21.9$\;$Mpc)\\ 
\noalign{\smallskip}
\hline
\noalign{\smallskip}
\end{tabular}

{\footnotesize
  Sources: 
 (1) Harris et~al. 1991; Secker \& Harris 1993; Forbes 1996
 (2) Ajhar et~al. 1994
 (3) Harris et~al. 1991; Ajhar et~al. 1994; Cohen 1988
 (4) van den Bergh et~al. 1985; Harris et~al. 1991; Cohen 1988;
     McLaughlin et~al. 1994; Whitmore et~al. 1995;
     Elson \& Santiago 1996a,b
 (5) Kissler et~al. 1994
 (6) Harris et~al. 1991.
  -- The values in parentheses in columns~2 and 3 give the number of
  independent determinations.}
\end{minipage}
\end{center}
\end{table}

   As seen in Table~\ref{tab:Virgo_GC} the GCLF leads to
\begin{displaymath}
  (m-M)_{\rm Virgo}=31.70\pm0.30,
\end{displaymath}
which is adopted in the following. For the adopted error see below.

   The question arises whether it is justified to apply $M^{\ast}$,
calibrated in two {\em spiral\/} galaxies, to the GCs of the
early-type galaxies in Table~\ref{tab:Virgo_GC}. A positive answer
within the errors is provided by the Leo group. Two early-type
galaxies in this group give a GCLF modulus of $(m-M)=30.08\pm0.29$
whereas the Cepheids in three spirals of the same group give
$(m-M)=30.20\pm0.12$ (Tammann \& Sandage 1999).

   The formation of GCs not being understood, there is no
theoretical reason why the value of $M^{\ast}$ should be universal. It
is worrisome that the {\em width\/} of the GCLF varies significantly
for different galaxies and that the two brightest galaxies in
Table~\ref{tab:Virgo_GC}, NGC\,4486 (M\,87) and NGC\,4472 (M\,49), as
well as NGC\,4552 (M\,89) have $m_{\rm V}^{\ast}$ $0\mag5$ {\em
  brighter\/} than the remaining four galaxies. Moreover, the color
$m_{\rm B}^{\ast} - m_{\rm V}^{\ast}=1.08$ of NGC\,4486 is
exceptionally red, and its GCLF seems to be bimodal. Finally it is
alarming that the mean GCLF modulus of seven early-type Fornax cluster
members is $0\mag54\pm0\mag15$ {\em smaller\/} than the secure cluster
distance from three blue SNe\,Ia, the latter value being also
supported by the relative distance from secondary distance indicators
between the early-type members of the Virgo and Fornax clusters
(Tammann \& Sandage 1999; cf. Fig.~\ref{fig:schematic_Ho_all}
below). In addition the GCLF distances of some individual field
galaxies are highly questionable. 

   Account of these problems is taken by assigning a relatively large
error to the GCLF distance of the Virgo cluster.

\section{The Virgo cluster distance from the  D\boldmath$_{n}-\sigma$ relation}
\label{sec:6}
It has been shown that the well known D$_{\rm n}-\sigma$ relation of
early-type galaxies (Dressler et~al. 1987) applies also to the bulges
of early-type spiral and S0 galaxies with surprisingly small scatter
(Dressler 1987). Here D$_{\rm n}$ is an isophotally defined galaxy
diameter (in arcsec), i.e. the diameter of a circle which encompasses a
mean surface brightness of 19.75 B\,mag arcsec$^{-2}$, and $\sigma$ is
the aperture-dependent, normalized velocity dispersion $\sigma$ (in
km\,s$^{-1}$) in the bulge. The isophotal values of D$_{\rm n}$ are,
of course, affected by front absorption $A_{\rm B}$. The corresponding
correction amounts to $\Delta \log D=0.32 A_{\rm B}$ (Lynden-Bell
et~al. 1988), which translates into a distance effect of $\Delta
(m-M)=1.6 A_{\rm B}$. The absorption has paradoxically thus a stronger
effect on this diameter distance than on a luminosity distance.

   From a sample of 26 S0--Sb Virgo cluster members Dressler (1987)
has determined 
\begin{equation}
  \label{equ:Dn_sigma}
  \log D_{\rm n}=1.333 \log \sigma - (1.572\pm0.014)
\end{equation}
with a scatter of $\sigma(\log D_{\rm n})=0.06$, corresponding to
$\sigma(m-M)=0\mag34$. The two deviating galaxies NGC\,4382 and
NGC\,4417 were excluded. 

   It is somewhat worrisome that equation~(\ref{equ:Dn_sigma}) is
based on only 24 galaxies, i.e. one third of the total Virgo
population of S0--Sb galaxies. This may invite selection
bias. However, the brighter half of the sample yields  the same
distance modulus as the fainter one to within
$0\mag07\pm0\mag14$. Moreover, the missing Virgo members are on
average fainter (smaller) at any given value of $\sigma$ than the
sample of 24, and they can make the constant term in
equation~(\ref{equ:Dn_sigma}) only more negative, which would in any
case {\em increase\/} the cluster distance.

   Yet for another reason the Virgo distance derived from
equation~(\ref{equ:Dn_sigma}) may be somewhat low. As discussed below
(Section~\ref{sec:8}) the Virgo cluster consists of two main
concentrations A and B, B being, if anything, slightly more
distant. The Virgo distance quoted throughout refers to the mean of
{\em all\/} cluster members and therefore depends on a fair
representation of A and B. In the case of
equation~(\ref{equ:Dn_sigma}) 21 galaxies lie in A and only 3 in B,
which is an overrepresentation of A even if one allows for the smaller
size of B.

   The Virgo cluster distance can be derived from
equation~(\ref{equ:Dn_sigma}) if the {\em linear\/} diameters D$_{\rm
  n}$ at given $\sigma$ are known. As calibrators have been used M\,31
and M\,81 (Dressler 1987; Sandage \& Tammann 1988) as well as the
bulge of our Galaxy (Terndrup 1988); all three galaxies have been
combined by Tammann (1988). The Galaxy, as a calibrator, may be of
somewhat later type than the S0--Sb sample, but also the Sbc galaxy
NGC\,4501 fits well equation~(\ref{equ:Dn_sigma}) (Dressler 1987). In
principle the Virgo cluster distance has to be known to match the
velocity dispersion $\sigma$ of the calibrators to the aperture size
($5'' \times 5''$) that was applied for Virgo. However, the remaining
mismatch should introduce only negligible systematic errors
(cf. Dressler 1987).
The derivation of the cluster distance is repeated in
Table~\ref{tab:Virgo_distance_D_sigma} with updated Cepheid distances
and absorption values for M\,31 and M\,81. The Table is
self-explanatory.

%
\begin{table}[t]
\footnotesize
\begin{center}
\caption{The local calibration of the D$_{\rm n}-\sigma$ relation of
  the bulges of early-type spirals and S0 galaxies}
\label{tab:Virgo_distance_D_sigma}
\begin{minipage}{\textwidth}
\begin{tabular}{lccccrc}
\noalign{\smallskip}
\hline
\noalign{\smallskip}
 Calibrator & $(m-M)^0$ & $\log$ D$_{\rm n}$ & $\sigma(\kms)$ & 
 $\log$\,D$_{\rm n}$\,(eq.\,\ref{equ:Dn_sigma}) & $\Delta(m-M)$ & $(m-M)^0$ \\
\noalign{\smallskip}
\hline
\noalign{\smallskip}
 Gal. Bulge & $14.46\pm.20^{1)}$ & $4.74\pm.04^{1)}$ & $124\pm9^{1)}$ &
           $1.22\pm0.06$ & $17.60\pm0.30$ & $32.06\pm0.36$ \\
 M\,31      & $24.44\pm.20^{2)}$ & $2.83\pm.04^{4)}$ & $150\pm5^{6)}$ &
           $1.33\pm0.04$ &  $7.50\pm0.20$ & $31.94\pm0.28$ \\
 M\,81      & $27.80\pm.20^{3)}$ & $2.15\pm.03^{5)}$ & $166\pm8^{6)}$ &
           $1.39\pm0.04$ &  $3.80\pm0.19$ & $31.60\pm0.28$ \\  
\noalign{\smallskip}
\hline
\noalign{\smallskip}
 \multicolumn{4}{l}{weighted mean:} &
 \multicolumn{3}{r}{$31.85\pm0.17$} \\
\noalign{\smallskip}
\hline
\noalign{\smallskip}
\end{tabular}
(1) Terndrup 1988
(2) Madore \& Freedman 1991
(3) Freedman \& Madore 1994
(4) Dressler 1987. D$_{\rm n}$ corrected for bulge absorption of
    $A_{\rm B}=0.33$; cf. text
(5) Dressler 1987. D$_{\rm n}$ corrected for front absorption of
    $A_{\rm B}=0.16$ (Freedman \& Madore 1994); cf. text
(6) Dressler 1987. 
\end{minipage}
\end{center}
\end{table}
 
  The method is so promising that one would hope for additional bulge
data of local calibrators as well as remaining Virgo cluster members
of the appropriate Hubble types. At present we adopt $(m-M)_{\rm
  Virgo}=31.85\pm0.17$ as the best D$_{\rm n}-\sigma$ distance from
intermediate-type galaxies.

   The classical D$_{\rm n}-\sigma$ relation of {\em early-type\/}
(E/S0) galaxies encounters the difficulty of lacking local
calibrators. The only way is to use the two early-type members of
the Leo group and to adopt the mean Cepheid distance of
$(m-M)=30.20\pm0.12$ of the three spirals in the same group. Faber
et~al. (1987) find the modulus difference between the Virgo cluster
and the Leo group to be $\Delta(m-M)=0.97\pm0.29$ from the D$_{\rm
  n}-\sigma$ relation. This value is somewhat suspicious because it is
significantly smaller than from four other relative distance
indicators (cf. Tammann \& Federspiel 1997). But taken the modulus
difference at face value, one obtains $(m-M)_{\rm Virgo}=31.17\pm0.31$
from E/S0 galaxies.

   The weighted mean distance modulus of the intermediate-type and
early-type Virgo cluster members becomes then
\begin{displaymath}
  (m-M)_{\rm Virgo}=31.70\pm0.15.
\end{displaymath}

\section{The Virgo cluster distance from Novae}
\label{sec:7}
Pritchet \& van~den Bergh (1987) found from six novae in Virgo
cluster ellipticals that they are $7\mag0\pm0\mag4$ more
distant than the {\em apparent\/} distance modulus of M\,31 of
$(m-M)_{\rm AB}=24.58\pm0.10$ from Cepheids
(Madore \& Freedman 1991) {\em and\/} Galactic novae
(Capaccioli et~al 1989). Livio (1997) found from a
  semi-theoretical analysis of the six Virgo novae $(m-M)_{\rm
  Virgo}=31.35\pm0.35$. 

A low-weight mean of $(m-M)_{\rm
  Virgo}=31.46\pm0.40 $ is adopted.

\section{Conclusions}
\label{sec:8}
The results of Sections~2\,-\,7 are compiled in
Table~\ref{tab:Virgo_distance_compilation}. The individual values lead
to a weighted mean distance modulus of $(m-M)_{\rm
  Virgo}=31.60\pm0.09$, corresponding to a distance of $r=20.9\pm0.9\;$Mpc.

%
\begin{table}[t]
\footnotesize
\begin{center}
\caption{Compilation of the different Virgo cluster moduli}
\label{tab:Virgo_distance_compilation}
\begin{tabular}{lllll}
\noalign{\smallskip}
\hline
\noalign{\smallskip}
  Method & $(m\!-\!M)_{\rm Virgo}$ & Type & Calibration & Source \\
\noalign{\smallskip}
\hline
\noalign{\smallskip}
 Cepheids             & $31.45 \pm 0.21$ & S     & $(m\!-\!M)_{\rm LMC}=18.50$
   & Tammann \& Sandage 1999 \\
 Supernovae Ia        & $31.55 \pm 0.20$ & S     & Cepheids 
   & Tammann \& Reindl 1999 \\ 
 Tully-Fisher         & $31.58 \pm 0.24$ & S     & Cepheids 
   & Federspiel et~al. 1998 \\
 Globular Clusters    & $31.70 \pm 0.30$ & E     & RR\,Lyr, Cepheids
   & Tammann \& Sandage 1999 \\
 D$_{\rm n} - \sigma$ & $31.70 \pm 0.15$ & E,\,S0,\,S & Galaxy, Cepheids
   & Dressler 1987 \\
 Novae                & $31.46 \pm 0.40$ & E     & M\,31 (Cepheids)
   & Pritchet \& van\,den Bergh$\;$1987 \\
\noalign{\smallskip}
\hline
\noalign{\smallskip}
 Mean: & $31.60\pm0.09$ & 
 \multicolumn{3}{l}{($\Rightarrow 20.9\pm0.9\;$Mpc)}\\
\noalign{\smallskip}
\hline
\noalign{\smallskip}
\end{tabular}
\end{center}
\end{table}

   It is remarkable that the individual distance determinations agree
to within their mean internal errors. The result gains additional
weight by the fact that it is based on spiral as well as early-type
cluster members. The zeropoint of the distance determination, as seen
in Table~\ref{tab:Virgo_distance_compilation}, rests mainly, but not
exclusively on Cepheids, and hence on the adopted LMC distance.

   The Virgo cluster shows clear subclustering. There are two major
clumpings A (with M\,87) and B (with M\,49) (Binggeli et~al. 1985,
1993) as well as a concentration around M\,86 (Schindler, Binggeli, \&
B{\"o}hringer 1999). There is rather strong evidence from the TF
method that cluster B is more distant than A by $\sim\!0\mag46\pm0.18$
(Federspiel et~al. 1998). However, at present it is save to include
{\em all\/} cluster members as defined by Binggeli et~al. (1993) and
to quote a single mean distance of the common gravitational well. 

   Future determinations of the Virgo cluster distance may include the
brightness of the tip of the red-giant branch (TRGB). A first
experiment is available (Harris et~al. 1998), and if a sufficient
number of cluster members will become available to beat the cluster
depth effect the method may become competitive. 

   It has been argued many times that the recession velocity of the
Virgo cluster is too small to yield the cosmic value of the Hubble
constant $H_0$. Indeed the observed mean cluster velocity of
$v_0=920\pm35\kms$, corrected to the centroid of the Local Group
(Binggeli et~al. 1993) and combined with the above cluster distance
would provide too low a value of $H_0=44$ [km\,s$^{-1}\;$Mpc$^{-1}$]
mainly due to the gravitational deceleration by the Virgo
complex. The Virgocentric pull has decelerated the Local Group's
recession velocity by $200-250\kms$ (Kraan-Korteweg 1986; Tammann \&
Sandage 1985; Jerjen \& Tammann 1993). $H_0$ becomes then rather 54
with some leeway as to remaining peculiar velocity effects.

   However the problem of the Virgo cluster velocity can be entirely
circumvented by using the distances of clusters out to $10\,000\kms$
{\em relative\/} to the Virgo cluster (Sandage \& Tammann 1990; Jerjen
\& Tammann 1993; Giovanelli 1997). The exquisite quality of these
relative distances is shown by their defining a Hubble line of slope
0.2 with very small scatter (Fig.~\ref{fig:Hubble_relative_distances}).

\def\floatwidth{0.665\textwidth} %
\begin{figure}[t]
\begin{center}
\centerline{\psfig{file=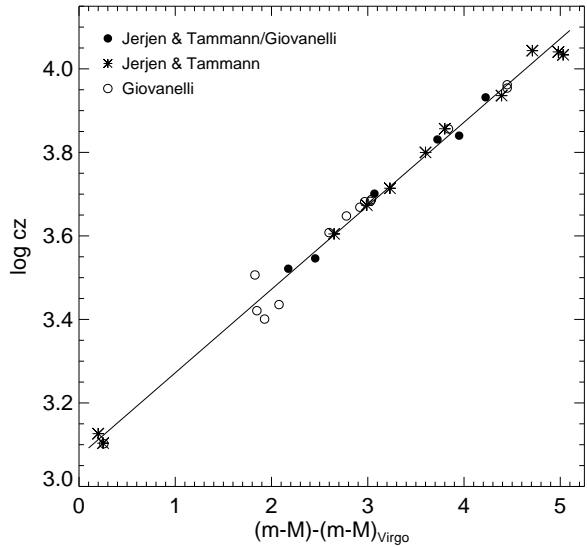,width=\floatwidth}}
\caption{Hubble diagram of 31 clusters with known relative
     distances. Asterisks are data from Jerjen \& Tammann (1993). Open
     circles are from Giovanelli (1997). Filled circles are the
     average of data from both sources. Velocities of $\ge3000\kms$ are
     corrected for a local CMB anisotropy of $620\kms$.}
  \label{fig:Hubble_relative_distances}
\end{center}
\end{figure}

   A linear regression through the points in
Fig.~\ref{fig:Hubble_relative_distances} with forced slope of 0.2,
corresponding to linear expansion, gives
\begin{equation}
  \label{equ:log_v_relative}
  \log v = 0.2\,[(m-M)-(m-M)_{\rm Virgo}] + (3.070\pm0.011).
\end{equation}
From this follows directly
\begin{equation}
  \label{equ:log_H_relative}
  \log H_0 = -0.2\,(m-M)_{\rm Virgo} + (8.070\pm0.011).
\end{equation}
Inserting the mean Virgo cluster modulus from
Table~\ref{tab:Virgo_distance_compilation} yields
\begin{displaymath}
  \label{equ:H0_Dn_sigma)}
  H_0 = 56\pm4,
\end{displaymath}
where the additional external error is generously estimated to be $\pm6$.

\begin{figure}[t]
\begin{center}
{\footnotesize \sf
\def\floatwidth{3.18cm}
\setlength{\unitlength}{1mm}
\def\xx{120}
\def\yy{151}
\begin{picture}(\xx,\yy)(0,0)
 \put(0,0){\line(\xx,0){\xx}}
 \put(0,0){\line(0,\yy){\yy}}
 \put(0,\yy){\line(\xx,0){\xx}}
 \put(\xx,0){\line(0,\yy){\yy}}
 \put(110,146){\makebox(0,0)[t]{\normalsize \bf
    H\boldmath$_0$ $\approx$ 58}}
 \put(53,146){\makebox(0,0)[t]{\normalsize \bf H\boldmath$_0$ $=$ 56$\pm$4}}
 \put(53,138.5){\vector(0,0){3.5}}
 \put(40.5,134){\fbox{\parbox{2.2cm}{\centering
    Clusters out to \\ 10\,000\kms}}}
 \put(36,91){\fbox{\psfig{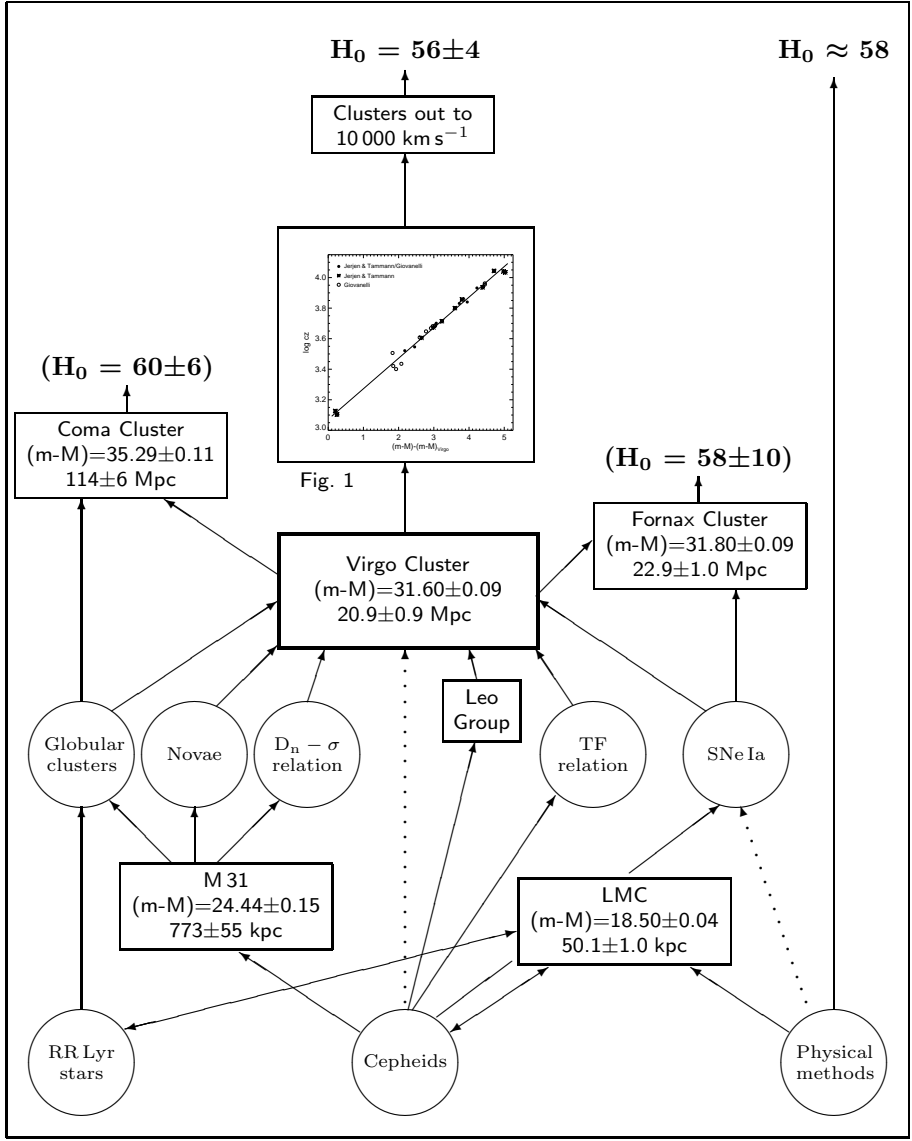}}}
 \put(39,86.5){Fig.~1}
 \put(53,121){\vector(0,0){10}}
 \put(53,80.3){\vector(0,0){9.5}}
 \put(36,74.9){\vector(-3,2){15}}
 \put(70.4,72){\vector(1,1){7.4}}
 \put(16,104){\makebox(0,0)[t]{\normalsize \bf
    (H\boldmath$_0$ $=$ 60$\pm$6)}}
 \put(16,96.5){\vector(0,0){3.5}}
 \put(1,90){\fbox{\parbox{2.6cm}{\centering
    Coma Cluster \\ (m-M)$=$35.29$\pm$0.11 \\
     114$\pm$6$\;$Mpc}}}
 \put(36,72){\fboxrule=1pt \fbox{\parbox{3.2cm}{\centering \vspace*{5pt}
    Virgo Cluster \\  (m-M)$=$31.60$\pm$0.09 \\
    20.9$\pm$0.9$\;$Mpc \vspace*{5pt}}}}
 \put(92,92){\makebox(0,0)[t]{\normalsize \bf (H\boldmath$_0$ $=$
   58$\pm$10)}}
 \put(92,84.5){\vector(0,0){3.5}}
 \put(78,78){\fbox{\parbox{2.6cm}{\centering
   Fornax Cluster \\  (m-M)$=$31.80$\pm$0.09 \\
   22.9$\pm$1.0$\;$Mpc}}}
 \put(58,56){\fbox{\parbox{0.8cm}{\centering
   Leo \\ Group}}}
 \put(62.5,61){\vector(-1,4){1}}
%
\put(14,56.7){\vector(3,2){22}}
\put(28,57.4){\vector(1,1){8}}
\put(40,58){\vector(1,3){2.3}}
\put(75.3,57.5){\vector(-2,3){5}}
\def\py{51}
 \put(10,58){\vector(0,0){27}}
 \put(10,58){\vector(0,0){27}}
\put(10,\py){\circle{40}}
\put(10,\py){\makebox(0,0)[c]{\parbox{1cm}{\centering \scriptsize
      Globular \\ clusters}}}
\put(25,\py){\circle{40}}
\put(25,\py){\makebox(0,0)[c]{\parbox{1cm}{\centering \scriptsize
      Novae}}}
\put(40,\py){\circle{40}}
\put(40,\py){\makebox(0,0)[c]{\parbox{1cm}{\centering \scriptsize
      D$_{\rm n}-\sigma$ \\ relation}}}
\put(25,36.5){\vector(0,0){7.5}}
\put(22,36.5){\vector(-1,1){8.3}}
\put(28,36.5){\vector(1,1){8.3}}
\put(78,\py){\circle{40}}
\put(78,\py){\makebox(0,0)[c]{\parbox{1cm}{\centering \scriptsize
      TF \\ relation}}}

\put(93,56.7){\vector(-3,2){22.3}}
\put(97,58){\vector(0,0){15.0}}
\put(97,\py){\circle{40}}
\put(97,\py){\makebox(0,0)[c]{\parbox{1cm}{\centering \scriptsize
      SNe\,Ia}}}
\def\py{30}
 \put(15,\py){\fbox{\parbox[c]{2.6cm}{\centering
   M\,31 \\ (m-M)$=$24.44$\pm$0.15 \\
     773$\pm$55$\;$kpc}}}
 \put(68,28){\fbox{\parbox{2.6cm}{\centering
   LMC \\ (m-M)$=$18.50$\pm$0.04 \\
     50.1$\pm$1.0$\;$kpc}}}
%
\def\py{10}
 \put(10,17){\vector(0,0){27}}
\put(15.8,14.4){\vector(4,1){52}}
\put(19.8,15.4){\vector(-4,-1){4}}
\put(10,\py){\circle{60}}
\put(10,\py){\makebox(0,0)[c]{\parbox{1cm}{\centering \scriptsize
      RR\,Lyr \\ stars}}}
%
\def\dx{53}
 \put(\dx,64){\vector(0,0){1}}
 \put(\dx,62){\circle*{0.5}}
 \put(\dx,60){\circle*{0.5}}
 \put(\dx,58){\circle*{0.5}}
 \put(\dx,56){\circle*{0.5}}
 \put(\dx,54){\circle*{0.5}}
 \put(\dx,52){\circle*{0.5}}
 \put(\dx,50){\circle*{0.5}}
 \put(\dx,48){\circle*{0.5}}
 \put(\dx,46){\circle*{0.5}}
 \put(\dx,44){\circle*{0.5}}
 \put(\dx,42){\circle*{0.5}}
 \put(\dx,40){\circle*{0.5}}
 \put(\dx,38){\circle*{0.5}}
 \put(\dx,36){\circle*{0.5}}
 \put(\dx,34){\circle*{0.5}}
 \put(\dx,32){\circle*{0.5}}
 \put(\dx,30){\circle*{0.5}}
 \put(\dx,28){\circle*{0.5}}
 \put(\dx,26){\circle*{0.5}}
 \put(\dx,24){\circle*{0.5}}
 \put(\dx,22){\circle*{0.5}}
 \put(\dx,20){\circle*{0.5}}
 \put(\dx,18){\circle*{0.5}}
\put(47,14){\vector(-3,2){16}}
\put(59,14){\vector(3,2){13}}
\put(60.5,15){\vector(-3,-2){1.5}}
\put(54,17){\vector(2,3){19}}
\put(53.4,17){\vector(1,4){8.9}}
\put(57.2,16){\line(4,3){10}}
\put(82.8,35.2){\vector(4,3){12}}

\put(\dx,\py){\circle{60}}
\put(\dx,\py){\makebox(0,0)[c]{\scriptsize Cepheids}}
 \put(98.6,41.1){\vector(-1,3){1}}
 \put(98.5,41.2){\circle*{0.5}}
 \put(99.1,39.4){\circle*{0.5}}
 \put(99.7,37.6){\circle*{0.5}}
 \put(100.3,35.8){\circle*{0.5}}
 \put(100.9,34.0){\circle*{0.5}}
 \put(101.5,32.2){\circle*{0.5}}
 \put(102.1,30.4){\circle*{0.5}}
 \put(102.7,28.6){\circle*{0.5}}
 \put(103.3,26.8){\circle*{0.5}}
 \put(103.9,25){\circle*{0.5}}
 \put(104.5,23.2){\circle*{0.5}}
 \put(105.1,21.4){\circle*{0.5}}
 \put(105.7,19.8){\circle*{0.5}}
 \put(106.3,18){\circle*{0.5}}
 \put(110,17){\vector(0,0){124}}
 \put(104,14){\vector(-3,2){13}}
\put(110,\py){\circle{60}}
\put(110,\py){\makebox(0,0)[c]{\parbox{1cm}{\centering \scriptsize
      Physical \\ methods}}}
%
\end{picture}
}
  \end{center}
  \caption{Schematical presentation of the distance determination of
    the Virgo cluster} 
  \label{fig:schematic_Ho_all}
\end{figure}

   The route to the Virgo cluster distance and to $H_0$ is
schematically summarized in Fig.~\ref{fig:schematic_Ho_all}. 

The figure also shows the distance modulus of the Fornax cluster. The
modulus holds strictly only for the early-type cluster members. The
Cepheid distance of the large {\em spiral\/} NGC\,1365 of $(m-M)=
31.35$ (Madore et~al. 1998) proves it to lie on the near side of
the cluster. Also other cluster spirals may be at relatively small
distances because a compilation of about 30 determinations by various
methods and authors of the relative distance between the Fornax and
Virgo clusters suggests that the Fornax spirals are nearer on average
than the Fornax E/S0 members by $0.35\pm0.10$ (Tammann \& Federspiel 1997).
In any case the Fornax cluster distance is not helpful for the
determination of $H_0$ because the observed mean cluster redshift of
$\sim\!1300\kms$ carries an uncertainty of $\sim\!20\%$ due to its
totally unknown peculiar motion.

   The Coma cluster distance in Figure~\ref{fig:schematic_Ho_all} is
not either very helpful for the determination of $H_0$. Its main
weight hinges on distances relative to the Virgo cluster and the
cluster contributes therefore only a limited amount of independent
evidence. Moreover, its observed mean velocity of $6900\kms$ may be
affected by local streaming velocities at the level of about 10\%. The
cluster lies probably outside the large local bubble which moves with
$630\kms$ toward the warm pole of the MWB (Smooth et~al. 1991), but it
seems plausible that it has a similarly large peculiar motion by its own.

   The value of $H_0\approx58$, also shown in
Fig.~\ref{fig:schematic_Ho_all}, is based on purely physical distance
determinations, i.e. gravitationally lensed quasars, the
Sunyaev-Zeldovich effect, and CMB fluctuations. The reader is referred
to the abstract by G. Theureau \& G.A. Tammann in these Conference
Proceedings where the original authors are quoted.

\section*{Acknowledgments}

G.A.T. and B.R. thank the Swiss National Science Foundation for
financial support.


\end{document}